\documentclass[aps,twocolumn,prd,showpacs,nofootinbib]{revtex4}
\usepackage{amsmath}
\usepackage{graphicx}
\usepackage{dcolumn}
\usepackage{bm}
\usepackage{amssymb}
\usepackage{latexsym}

\def\be{\begin{equation}}
\def\ee{\end{equation}}
\def\ba{\begin{eqnarray}}
\def\ea{\end{eqnarray}}

\begin{document}

\title{Proliferation in Cycle  }

\author{Yun-Song Piao}

\affiliation{College of Physical Sciences, Graduate School of
Chinese Academy of Sciences, Beijing 100049, China}

\begin{abstract}

In the contracting phase with $w\simeq 0$, the scale invariant
spectrum of curvature perturbation is given by the increasing mode
of metric perturbation. In this paper, it is found that if the
contracting phase with $w\simeq 0$ is included in each cycle of a
cycle universe, since the metric perturbation is amplified on
super horizon scale cycle by cycle, after each cycle the universe
will be inevitably separated into many parts independent of one
another, each of which corresponds to a new universe and evolves
up to next cycle, and then is separated again. In this sense, a
cyclic multiverse scenario is actually presented, in which the
universe proliferates cycle by cycle. We estimate the number of
new universes proliferated in each cycle, and discuss the
implications of this result.

\end{abstract}

%\pacs{98.80.Cq}

\maketitle

Recently, it has been found that the contraction of universe with
the state parameter $w\simeq 0$ can lead to a nearly scale
invariant spectrum of primordial perturbation \cite{Wands99, FB},
also see \cite{PP},\cite{Wands09},\cite{Cai0810} and earlier
\cite{S} for tensor perturbation, which can be responsible for the
seed of observable universe after bounce. This spectrum of
curvature perturbation is actually given by the increasing mode of
metric perturbation, in which the increasing mode of metric
perturbation is inherited by the constant mode of curvature
perturbation in its $k^2$ order. Thus this equals that it is
amplified on super horizon scale, up to bounce. The amplitude that
it is amplified to is determined by the bounce scale, and thus in
some sense the efolding number before the end of contracting
phase.

%Historically, the idea of cyclic universe, in ,
%has been still interesting .
Recently, the cyclic scenario, in which the universe experiences
the periodic sequence of contractions and expansions\cite{Tolman},
has been rewaked \cite{STS}, since it brings the different
insights for the origin of observable universe. There has been
lots of studies for oscillating or cyclic universe
\cite{LS},\cite{Xiong},\cite{BD},\cite{KSS},\cite{Piao04},\cite{Lidsey04},
\cite{BF},\cite{CB},\cite{Xin},\cite{Biswas}, also \cite{NB} for a
review. In principle, it is interesting to include our observable
universe in such a cyclic universe. The nearly scale invariant
spectrum of primordial perturbation is required for the structures
in our universe. Thus in this sense, we should consider the effect
of perturbation on a cyclic universe. In another viewpoint, it is
generally expected that the background of cycle universe should be
homogenous all along during all cycles. However, since the
perturbation is increased during the contraction of each cycle, it
is required to check whether such homogeneousity can be still
preserved as expected.

The contracting phase with $w\simeq 0$ is suitable for the
observable universe after bounce. In principle, it can be
considered to involve in a cyclic scenario. We, in this paper,
will explore that if the contracting phase with $w\simeq 0$ is
included in each cycle of a cycle universe, what occurs. We find
that the amplitude of perturbation modes that are generated in
previous cycle and still stay on super horizon scale will be
inevitably amplified to about order one at about beginning time of
following cycle, which will render the different parts of global
universe in this cycle evolve not anymore synchronously and
decouple each other. This implies that after each cycle the
universe will be separated into many parts independent of one
another, each of which corresponds to a new universe and evolves
up to next cycle, and then is separated again \footnote{The
similar phenomena has been discussed in Ref. \cite{Erick} for
cyclic universe \cite{STS}, in which the increasing mode of metric
perturbation is inherited by the constant mode of curvature
perturbation in leading order, which is controversial
\cite{FB},\cite{DH},\cite{DV},\cite{TBF},\cite{GKS},\cite{Piao0403},\cite{BV},\cite{ABB},\cite{Cai11},
and is different from that discussed here, in which the increasing
mode of metric perturbation is inherited by the constant mode of
curvature perturbation in its $k^2$ order, which certainly occurs
for the bounce connecting the contracting and expanding phases,
e.g. \cite{AW}.}. This result shows that the universe proliferates
cycle by cycle, which looks like a cyclic multiverse. We will
calculate the number of new universes proliferated in each cycle,
and investigate how it is affected by other factors. How this
cyclic multiverse scenario incorporates the second law of
thermodynamics is also discussed.

We will regard the beginning of the contracting phase as the
beginning of a cycle, in each cycle the universe will experience
the contraction, bounce, and expansion, successively, and then
arrive at the turnaround, which signals the end of a cycle. For
generality of the result, we will not involve the building of some
special cyclic models, however, which is actually not difficult.
We will begin with the review of the perturbation spectrum
generated during the contraction with $w\simeq 0$ \cite{Wands99,
FB}. In principle, the contraction with $w\simeq 0$ can be easily
implemented by introducing a scalar field $\varphi$ with the
suitable exponent potential, which leads a scale solution for
$\varphi$, see e.g.\cite{FB}. In this case, the motion equation of
curvature perturbation is \be u_k^{\prime\prime}
+\left(k^2-{z^{\prime\prime}\over z}\right) u_k = 0 ,\label{uk}\ee
where $u_k$ is related to the curvature perturbation $\zeta$ by
$u_k \equiv z\zeta_k$ and the prime denotes the derivative with
respect to the conformal time $\eta$, and $z={{\varphi^\prime}/
h}$, where $h$ is the Hubble parameter. For $w\simeq 0$, we have
$a\sim \eta^2$ and $h\sim 1/\eta^3$. The scale solution of
evolution of $\varphi$ gives ${\dot \varphi}/h$ constant, which
leads $z\sim \eta^2$. Thus we have ${z^{\prime\prime}\over z} \sim
{2\over \eta^2}$, which is actually the same as that for
inflation, in which $a\sim 1/\eta$ leads $z\sim 1/\eta$ and thus
${z^{\prime\prime}\over z} \sim {2\over \eta^2}$. This gives the
spectrum $n_s\simeq 1$ is scale invariant, and the amplitude of
perturbation is \be {\cal P}_{\zeta}^{1/2}\simeq k^{3/2}|{u_k\over
z}|\simeq {h_b\over m_p}, \label{p1}\ee where the factor of order
one has been neglected and $h_b$ is determined by the energy scale
$\rho_b$ of field around the bounce point, $h_b\simeq
{\sqrt{\rho_b}\over m_p}$. The amplitude of this perturbation
spectrum is actually leaded by the increasing mode of metric
perturbation $\Phi$ \cite{FB}
%\footnote{The general solutions of
%metric perturbation $\Phi$ are $\Phi \simeq \Phi_C+ \Phi_S{h\over
%a}$, where for $w\simeq 0$, $\Phi_C\sim k^{3\over 2}$ is the
%constant mode, and $\Phi_S\sim 1/k^{{7\over 2}}$ is the increasing
%mode, which dominates $\Phi$ during contraction. Both are only $k$
%dependent functions. The solution of $\zeta$ is about $ {\Phi_C}+
%{\Phi_S}f(\eta)k^2$, where $f(\eta)\simeq {h\over m_p}$ for
%$w\simeq 0$. Thus the $\Phi_S$ mode enters into $\zeta$ in its
%$k^2$ order. In general, $\zeta$ and $\Phi$ are continuous during
%bounce. Thus with Refs. \cite{HV, DM}, $\Phi_C^{+}\simeq
%\Phi_C^{-}+ \Phi_S^{-} fk^2$ can be obtained \cite{FB}, where
%$\Phi_C^{-}$ mode is subdominated and can be neglected, and the
%superscripts '$+$' and '$-$' denote the quantities after and
%before bounce, respectively. Thus the increasing mode $\Phi_S^{-}$
%before bounce is inherited by the constant mode $\Phi_C^{+}$ in
%its $k^2$ order after bounce. While the $\Phi_S^{+}$ mode after
%bounce is the decaying mode, which is generally negligible. Thus
%the spectrum of perturbation after bounce is \be {\cal
%P}_{\zeta}^{1/2}\simeq k^{3/2}|\Phi_C^+(\eta_b)|\simeq {h_b\over
%m_p}. \label{pp2}\ee}
,
%in which it was showed that generally the increasing
%mode of metric perturbation can be inherited by the constant mode
%of curvature perturbation in its
%$k^2$ order,
see also Ref. \cite{Cai0810} for more details. This means the
amplitude of perturbation begins to be amplified gradually on
super horizon scale, up to the bounce, since the perturbation mode
leaves the horizon. This will be formulized in term of the
efolding number in the following.

%When
%${h_b\over m_p}\sim 10^{-5}$, the universe after bounce can be
%that like ours.

%The universe can stop expanding and begin to contract after some
%time. This can be implemented simply by introduce the negative
%potential.
The efolding number of mode with some wavelength $\sim 1/k_*$,
which leaves the horizon before the end of contracting phase, is
defined by \be {\cal N} \equiv \ln{({a_b h_b\over a_* h_*})}
,\label{ne}\ee which is actually not the efolding number of scale
factor, but is that of primordial perturbation, and $k_b$ is the
last mode to be generated. We generally have ${\cal N}\sim 50$,
which is required by observable cosmology. In term of $a\sim
\eta^2$ and $h\sim 1/\eta^3$, we have $a\sim 1/h^{2/3}$. Thus the
efolding number during contraction can be deduced, which is ${\cal
N}_{con}=\ln{({h_b\over h_*})^{1/3}}$. We substitute it into
Eq.(\ref{p1}) to cancel $h_b$, and obtain \be {\cal
P}_{\zeta}^{1/2}\simeq {h_*\over m_p}e^{3{\cal N}_{con}}.
\label{p2}\ee This result implies that after the perturbation
denoted by $k_*$ leaves the horizon, the multiple that its
amplitude is amplified is given by the efolding number, while
initially, i.e. it just leaves the horizon, its amplitude is
approximately given by ${h_*\over m_p}$. This shows that if we
regard the initial amplitude of perturbation outside horizon as
${h_*\over m_p}$, then hereafter during contraction the multiple
that it is amplified on super horizon scale will be determined by
the efolding number before the end of contracting phase, which is
the result required for following arguments.

The entropy is inevitably increased during each cycle, which is
the requirement of the second law of thermodynamics, thus it is
expected that the maximal values which the scale factor can arrive
at in ordinal cycles should larger and larger. In this case, there
are certainly some modes that leave the horizon during the
contraction of each cycle but can not reenter into the horizon
during the expansion of corresponding cycle. These modes will
destine to stay on super horizon scale all along up to next cycle,
see the upper panel of Fig.1. We will regard the previous and
current cycles as $i$ and $j$ cycles for convenience,
respectively, in which $j=i+1$. The perturbation during the
contraction after turnaround satisfies Eq.(\ref{uk}). Thus if the
perturbation mode is initially deep in the horizon of current
cycle, its amplitude after bounce in current cycle will be given
by Eq.(\ref{p1}), i.e. \be {\cal P}_{\zeta (i)}^{1/2}\simeq
{h_b\over m_p}, \label{pi}\ee where the subscript $i$ denotes the
amplitude of perturbation after bounce in $i$ cycle, which is that
leaves the horizon during the contraction of $i$ cycle and
reenters into the horizon during the expansion of corresponding
cycle, thus same meaning for ${\cal P}_{\zeta (j)}^{1/2}$. Here,
the amplitude calculated is that of perturbation induced by the
quantum fluctuation of background field in current cycle. However,
in general it will be interfered by those perturbations that enter
into the horizon during the expansion of previous cycle. Thus the
actual amplitude can be larger, dependent of the expansion
behavior and matter contents of previous cycle, and the spectrum
is also not scale invariant any more. In this case, if we hope
that some of universes after proliferation could be same as ours,
it seems a period of dark energy in previous cycle or inflation
after bounce in current cycle is required, since it helps to push
those baneful modes to outside of our observable universe. This
will be discussed in details in coming works.

The amplitudes of those modes initially on super horizon scale at
current cycle should be calculated by using Eq.(\ref{uk}) again,
however, in which instead the initial condition is given by that
at the time of turnaround of the previous cycle. This value is
${h_b\over m_p}$, since the amplitude of perturbation on super
horizon scale after the bounce of previous cycle is dominated by
the constant mode, and thus is unchanged, which will be expected
to keep all along up to the turnaround of this cycle. However,
after the turnaround, the modes on super horizon scale will be not
unchanged any more. In term of the analogy with Eq.(\ref{p2}), in
$j$ cycle the amplitude of perturbation mode all along on super
horizon scale is given by \be {\cal P}_{\zeta
(j)}^{1/2}|_{i}\simeq {h_b\over m_p}e^{3{\cal N}_{con}}\simeq
{\cal P}_{\zeta (i)}^{1/2}e^{3{\cal N}_{con}}, \label{p3}\ee where
Eq.(\ref{pi}) has been applied, and ${\cal P}_{\zeta
(j)}^{1/2}|_{i}$ denotes the amplitude of perturbation, which
leaves the horizon during the contraction of $i$ cycle and
reenters into the horizon in the expanding phase of following $j$
cycle. Thus we see that in $j$ cycle it will be further amplified
with the proceeding of contracting phase.

The amplitude of perturbation responsible for seeding large scale
structure of observable universe is ${\cal P}_{\zeta
(i)}^{1/2}\sim 10^{-5}$. Thus if ${\cal P}_{\zeta (i)}^{1/2}\sim
10^{-5}$ is required in previous cycle, we can see when ${\cal
N}\simeq {5\over 3}\ln{10}\sim 3$, ${\cal P}_{\zeta (j
)}^{1/2}|_i\sim 1$ in $j$ cycle, where it should be noticed that
when ${\cal P}_{\zeta (j)}^{1/2}|_i$ approaches 1, the enhancement
of nonlinear effect will make the required $\cal N$ less. Thus in
fact this means that nearly at the beginning time of $j$ cycle,
the modes on super horizon scale will have the amplitude be about
order one. This will lead to ${\delta \rho\over \rho} \sim 1$ on
corresponding super horizon scale at this time. In this case, it
is obviously impossible that the different regions of global
universe will evolve synchronously, even if it is synchronous in
previous cycle. This indicates that the global universe at the
beginning time of this cycle will be separated into many different
parts, each of which will evolve independently of one another, up
to bounce. While inside any given part, all perturbation modes
origin from the interior of horizon, which is causally
correlative.
%their amplitudes are
%certainly initially quite small and can be expected to amplify to
%${\cal P}_{\zeta (j)}^{1/2}\simeq {h_b\over m_p}$ before the end
%of bounce in corresponding cycle.
Thus in this sense, each of such parts actually corresponds to a
new universe.
%which might have the
%scale invariant spectrum of perturbation and the suitable value of
%perturbation amplitude, and thus the structures like ours.
%Thus the universe will proliferate at the beginning time of each
%cycle.

In principle, each of these new universes will experience the
contraction, bounce and expansion, hereafter all or some of them
will enter into next cycle and proliferate again, and then the
above course is repeated again. This means the proliferation will
inevitably occur cycle by cycle. Thus we can have a cyclic
multiverse scenario. In this cyclic multiverse, the experience of
each universe after proliferation is generally not expected to be
synchronous. Thus when some universe are in a period of matter
domination, it is possible that there are many other universes
which are in the period of contraction or bounce or others. There
is also the proliferation of global universe in chaotic eternal
inflation \cite{83,86}, in which it is induced by the large
quantum fluctuation of inflaton field in its horizon scale, which
occurs efold by efold. Here, however, the proliferation is induced
by the cyclical amplification of perturbation on super horizon
scale, which is in classical sense, thus it occurs cycle by cycle.

The number of new universes proliferated at the beginning time of
each cycle is   \footnote{This value might be a conservative
estimate, since it is possible that those modes that enter into
the horizon during the expansion of $i$ cycle and then leave it
during the contraction of $j$ cycle may be also amplified to about
order one. In this case, the actually value will be larger. We
thank R. Brandenberger for talking this point to us. For this
case, the result given here can be regarded as a lower bound.
%it is quite possible that because of the effect of
%cycle, some peculiarities of spectrum can be expected to appear,
%and imprinted in the CMB of corresponding observable universe.
%is not certain that the spectrum
%of perturbation is not affected by its lingering inside the
%horizon in each cycle.
}\be { N}\simeq ({k_{j}\over k_{i}})^3=({a_{j}h_{j}\over
a_{i}h_{i}})^3, \label{nm}\ee where $k_i$ and $k_j$ denote the
modes with maximal wavelengths leaving the horizon at $i$ cycle
and at following $j$ cycle, respectively, and thus $1/k_i$ and
$1/k_j$ are the corresponding wavelengths, respectively. In this
sense, $a_{i,j}$ correspond to the magnitudes of scale factors at
the beginning time in $i$ and $j$ cycle, respectively, which
actually equals to the maximal magnitudes which the scale factors
expand to in previous cycle. $h_{i,j}$ are the turnaround scales
at corresponding cycles. The turnaround scale is generally same
for each cycle. Thus we have ${N}\simeq ({a_j\over a_i})^3$. We
can see that only when $a_j>a_i$ is there the appearing of many
new universes, while when $a_j=a_i$, ${ N}=1$. This means for
cyclic universe with equal cycles, the global universe in previous
cycle will not be separated into many new universes in current
cycle. The reason is simple, because there are not the
perturbation modes staying on super horizon scale all along up to
next cycle, see the lower left panel of Fig.2. It can be noticed
that after the proliferation in $j$ cycle the scale factor of each
new universe is approximately $a_i$. Thus for each of them, it can
be expected that what happened in $i$ cycle will be repeated,
which looks like that plotted in the lower panel of Fig.1.
However, it should be reminded that this is only an ideal case,
since, as has been mentioned, it is possible that those modes that
enter into the horizon during the expansion of $i$ cycle and then
leave it during the contraction of $j$ cycle may be also amplified
to about order one. In this case, it seems the universe will be
split into smaller and smaller parts, which will ultimately end
the cycle. This in some sense indicates that in cyclic universe a
gradually growing cycle is significant for the continuance of
cycle.

\begin{figure}[t]
\begin{center}
\includegraphics[width=7.5cm]{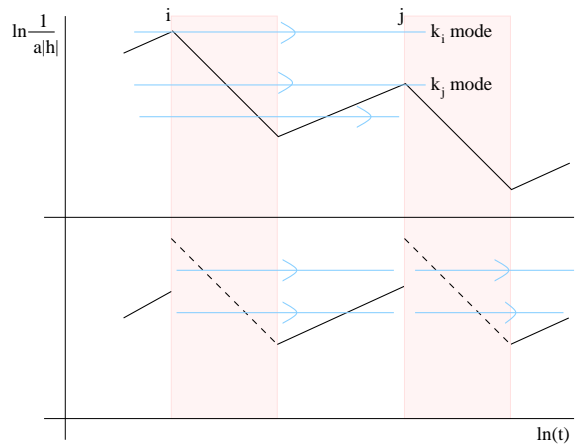}
\caption{ The sketch of $\ln{({1\over a|h|})}$ with respect to the
time. The blue lines denote the evolutions of perturbation modes
and the shade regions denote the contracting phases. The upper
panel shows the evolutions of perturbation modes in $i$ and
following $j$ cycles for the case that there is not the
proliferation in each cycle. There are certainly some modes that
leaves the horizon in $i$ cycle but can not enter into the horizon
during the expansion of corresponding cycle. These modes will
destine to stay on super horizon scale all along up to $j$ cycle.
The amplitude of these perturbations will be amplified to about
order one at about beginning time of $j$ cycle. This leads the
universe separated into many independent new universes, each of
which has the initial scale factor equal to that in previous
cycle. Thus for each universe in cycle, it seems that the
evolution of $\ln{({1\over a|h|})}$ with the time actually likes
the lower panel.
%Here Fig.1 and following Fig.2 are plotted in
%term of $\ln{({1\over ah})}\sim (1-p)\ln{t}$, in which $a\sim t^p$
%and $p$ is constant for corresponding evolution.
In principle, at the bounce and turnaround points, $h=0$, thus
there is a divergence for ${1\over a|h|}$, which is not plotted
here and Fig.2. However, the discussions are not affected by this
neglect. }
\end{center}
\end{figure}

%The maximal magnitude of scale factor in the expansion of each
%cycle is generally larger and larger because of the increase of
%the entropy in each cycle. For each expansion, in general the
%entropy $S\simeq a^3T^3$,
We can assume that after the bounce of each cycle the universe
enters into the expansion dominated by radiation and then matter.
Thus we have the radiation entropy $S\simeq a^3T^3$, which can be
enhanced by the decay of some relic massive particles, where $T$
is the corresponding temperature. For the period of matter
domination, $S$ corresponds to the CMB entropy. Though there are
the increase of super horizon perturbations, it can be showed that
the entropy in scalar and tensor perturbations is smaller than
that of CMB \cite{Brandenberger93},\cite{Brandenberger931}. Thus
around the turnround we have $S_{i,j}\simeq a^3_{i,j}T^3_{i,j}$,
where $S_i$ and $S_j$ denote the entropy at the beginning time in
$i$ and $j$ cycles, respectively. In this sense, $S_j$ is actually
the maximal value that the radiation entropy in $i$ cycle can
increase to. $T^3_{i,j}$ can be related to the turnaround scale
$h_{i,j}$ by $h_{i,j}\simeq {T^2_{i,j}\over m_p}$. Thus combining
these results with Eq.(\ref{nm}), then taking $h_j\simeq h_i$, \be
{N}\simeq {S_j\over S_i} \label{ns}\ee can be obtained, which
shows that the number of new universes proliferated in $j$ cycle
is determined by the net increasing amount of radiation entropy in
$i$ cycle.

%In general, it is possible that there will be lots of black hole
%produced in each cycle. Thus in principle, the entropy of black
%holes is dominated in the contribution to the entropy of
%observable universe. However, here only the CMB entropy is
%considered, since it is straightly related to the calculations
%here.

This result indicates that after the proliferation in each cycle,
each new universe in this cycle actually has the entropy equal to
$S_i$. This can be thought as if there is a net increase of the
entropy during some cycle, then in following cycle these net
entropy will be assigned to each new universes proliferated such
that the entropy of each of them is equal to that at the beginning
time of previous cycle. In this sense, it seems the problem of
entropy increase suffering the cyclic universe may be alleviated,
since if one begins with any given point in cyclic multiverse and
look along cycles, he will find in himself observable universe
that though the total entropy increases in each cycle, there is
not net increase of the entropy from one cycle to next. Here the
total entropy is the sum of entropy of all universes at some
spacelike hypersurface, which is obviously increased all along,
due to the second law of thermodynamics. In some sense, it seems
that here the second law of thermodynamics is incorporated in such
a fashion in which the increase of total entropy is explained as
the increase of the number of universes.

The inflation can be imagined to occur in some cycles. However,
regardless whether it occurs before turnaround, see the upper
right panel of Fig.2, or after bounce, see the upper left panel of
Fig.2, there will be more proliferated parts in next cycle, i.e.
new universes. The reason is a nearly exponent expansion makes the
scale factor in corresponding cycle larger, thus in term of
Eq.(\ref{nm}), for the fixed turnaround scale, $N$ will be larger.
From Eq.(\ref{nm}), we can obtain \be N\simeq ({a_j\over
a_i})^3\simeq ({a_j\over {a}_{j_*}})^3 e^{3{\cal
N}_{inf}},\label{nn}\ee where ${a}_{j_*}$ denotes the maximal
value that the scale factor can reach if there is not such an
inflation, and ${\cal N}_{inf}$ is the efolding number of
inflation, which may be given by taking the Hubble parameter
constant in Eq.(\ref{ne}). Thus with the increase of the efolding
number of inflation, the number of new universes proliferated in
following cycle will increase exponentially. This result is
natural, since it is generally expected that during the reheating
after inflation there will be lots of entropy relaxed, and the
ratio of the entropy after reheating to that before inflation is
approximately $e^{3{\cal N}}$, thus in this sense it is actually
the enhancement of entropy before and after inflation that leads
the increase of the number of new universes proliferated in
following cycle, which is consistent with Eq.(\ref{ns}). The
inflation after bounce was firstly studied in Refs. \cite{PFZ,
PTZ}, in which the imprint of bounce on CMB has been showed. It
can be noticed that if the inflation occurs before the turnaround
in some cycle, it actually corresponds to the period of dark
energy domination in previous cycle, see the upper right panel of
Fig.2.

It has been showed that each of new universes proliferated in each
cycle will evolve with the initial entropy $S_i$ and the initial
scale factor $a_i$ up to next cycle, and then proliferate in this
cycle, and again each of new universes still has the initial
entropy $S_i$ and the initial scale factor $a_i$ for following
evolution. Thus in this cyclic multiverse scenario the cycle is
actually eternal. However, this eternity seems not for past, an
initial condition might still be required. The reason is that the
number of universes increases with each cycle by a finite factor,
which inevitably leads that in the past the number of universes
reduces to one, thus there is again a big bang singularity. The
principal motivation for cyclicity is to avoid the big bang
singularity. This seems be lost here. However, whether the cyclic
universe avoids the big bang singularity is still a disputed
issue. In principle, if we consider the second law of
thermodynamics, since the entropy is increased cycle by cycle, the
length of cycle must continuously increase by a finite factor
cycle by cycle. Thus if we back along cycles, we will certainly
find a less and less length of cycle, up to `0' at a finite time.
In this sense, the cyclic universe actually dose not avoid the big
bang singularity. Here, the proliferation of universe cycle by
cycle is based on the increase of the length of cycles, and thus
the increase of entropy. In some sense, it equals to that here the
increase of the length of cycles in usual cyclic universe is
transferred to the increase of the number of universes. Thus if we
think that in a cyclic universe with the increase of entropy, the
problem of singularity remains, then it is same here. In
principle, the origin of this initial single universe needs to be
explained. However, of course, also it may be the one eternally
existing in the past, for example, origining from a steady state
background, like discussions in Ref. \cite{Biswas}.

The initial magnitude of scale factor of this initial universe
determines that of scale factor of each universe in multiverse,
for example, if we find the initial magnitude of scale factor is
$a_i$ in $i$ cycle, we will have $a_i$ for this initial single
universe. Thus in order to have an observable universe like ours
in corresponding cycle, it must be large. This seems to add a
requirement for initial condition. However, this can be relaxed as
follows. In general, the turnaround scale might be not same for
each cycle, for example, if $h_j<h_i$, it will be possible that
the initial magnitude $a_j$ of scale factor of the new universes
proliferated is larger than that of previous cycle, see the lower
right panel of Fig.2, since $a$ will continue to expand from the
scale $h_{j^{\prime}}=h_i$ to $h_{j}$, where $h_{j^{\prime}}$ is
the scale equal to the turnaround scale in previous cycle. In this
case, in term of Eq.(\ref{nm}), $N_{(h_j<h_i)}$ proliferated will
be less. This can be estimated by assuming that the evolution
between $h_{j^{\prime}}$ and $h_{j}$ is dominated by matter, \be
N_{(h_j<h_i)}\simeq ({h_j\over h_{j^{\prime}}})^{1\over 3}N \simeq
({h_j\over h_{i}})^{1\over 3}N, \label{nhh}\ee where $a\sim
1/h^{2/3}$ for the period of matter domination has been applied.

%Thus after the proliferation in following cycle, each universe has
%the initial magnitude of scale factor obviously larger than that
%in previous cycle.

%In this case, if the turnaround scale in ordinal cycle is
%gradually decreased, in principle the initial magnitude of scale
%factor in ordinal cycle is distinctly increased.

When the turnaround scale in some cycle is enough low, for example
$h_j\rightarrow 0$, it seems that in following cycle the initial
magnitude of scale factor of each universe can be infinite large.
However, the case is not so. When $h_j\rightarrow 0$,
$1/(a_jh_j)\sim 1/h_j^{1/3}$ is large so that $1/(a_jh_j)\gtrsim
1/(a_ih_i)$. In this case the modes that are generated in previous
cycle but can not enter into the horizon during the expansion of
corresponding cycle will be inevitable to enter into the horizon
of present universe, while these modes actually have amplitude
${\delta\rho\over \rho}\sim 1$. When these metric perturbations
enter into the horizon, they will certainly induce the
fluctuations of energy density on the corresponding horizon scale,
which will render the corresponding regions gravitational
collapse. This corresponds to set an upper limit for the initial
magnitude of scale factor of each universe proliferated in
following cycle. This result shows that the information before two
cycles is inaccessible to the observers in observable universe of
any given cycle. Thus if we are in some cycle of this cyclic
multiverse, we can at most see the modes produced in previous
cycle and thus information.

\begin{figure}[t]
\begin{center}
\includegraphics[width=7.5cm]{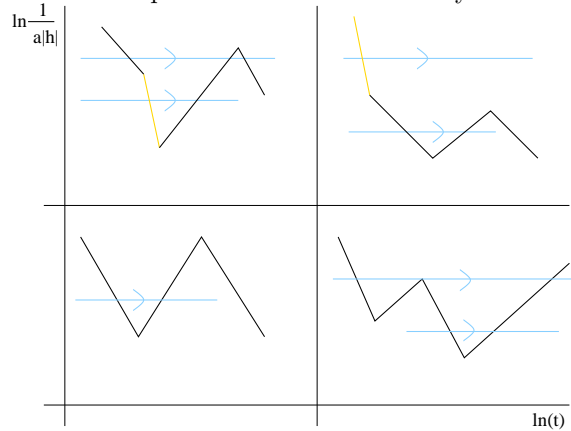}
\caption{ The sketch of $\ln{({1\over a|h|})}$ with respect to the
time for four different evolutions. The blue lines denote the
evolutions of perturbation modes and the yellow lines denotes the
inflation stages. The upper left panel is that there is an
inflation phase after bounce. The upper right panel is that there
is an inflation phase before turnaround, which actually
corresponds to the period of dark energy domination in previous
cycle. The lower left panel is that of cyclic universe with equal
cycles, in which in any given cycle there are not the super
horizon scale modes generated in previous cycle. The lower right
panel is that the turnaround scale in some cycle is lower than
that in previous cycle, thus the modes that are generated in
previous cycles but can not enter into the horizon during the
expansion of corresponding cycle can enter into the horizon of
this cycle.  }
\end{center}
\end{figure}

However, it can be noticed that this outcome is actually dependent
of the bounce scale. Here the bounce scale is high, thus the
amplitude after bounce in each cycle is large, which will be
inevitably amplified to order one in following cycle. However, if
the bounce scale is enough low, it is also possible that the
amplitude of perturbation leaving the horizon in some cycle can
not reach order one till several cycles. This will lead some
possible and interesting observations, which will be explored in
coming works.

%In this sense, it seems difficult to confirm this multiverse
%scenario by observations. However,

In conclusion, it is found that if the contracting phase with
$w\simeq 0$ is included in each cycle of a cycle universe, after
each cycle the universe will be inevitably separated into many
parts independent of one another, each of which corresponds to a
new universe and evolve up to next cycle, and then is separated
again. Thus a cyclic multiverse scenario is actually presented, in
which the universe proliferates cycle by cycle. This scenario are
leaded by the amplification of metric perturbation on super
horizon scale cycle by cycle, which can be general, since for the
contracting phase with $w\simeq 0$ the increasing mode of metric
perturbation is inherited by the constant mode of curvature
perturbation in its $k^2$ order is universal. We estimate the
number of new universes proliferated in each cycle, which, for
same turnaround scale, is determined approximately by the net
increasing amount of radiation entropy in previous cycle.
%In this sense, the increase of
%entropy cycle by cycle in usual cyclic universe is replaced with
%the multiple of number of new universes in each cycle.
This in some sense incorporates the second law of thermodynamics
in such a fashion in which the increase of total entropy is
explained as the increase of the number of new universes.

We have showed that the global configuration of cyclic universe is
more complex than expected ever, which actually shows itself a
cyclic multiverse. Though the arguments given here seems slightly
ideal, it might have captured some essentials of full answer. It
can be noticed that in general the background evolution of
contracting phase with $w\simeq 0$ is not an attractor. The
relevant discussions, also involving the landscape, is being
ordered.

%This work indicates the global configuration of cyclic universe
%might be more interesting and also complex than expected. Though
%this work seems slightly idealist, it might have captured some
%essentials of full answer, which is interesting for further
%studies.

%It should be mentioned that in order to have an estimate for the
%probability we identify the problem be effectively one
%dimensional. Thus it is inevitable that the result obtained is
%slightly rough. %In principle, Eq.(\ref{allpequ}) needs to be
%solved exactly.
%which, however, might have captured some essentials of full
%answer.

%the discussions here

%Though this work seems slightly idealist, This indicates the
%configuration of cyclic universe might be more complex than
%expected, which might be interesting for further studies.

%The entropy in each cycle can be produced by the decay of relic
%massive particles and during the reheating after inflation.

%Thus we have ${z^{\prime\prime}\over z}\simeq {2\over \eta^2}$,
%which means the spectrum is scale invariant. The amplitude of
%spectrum is given by \be {\cal P}_{\zeta}^{1/2}\simeq {h_b\over
%m_p}, \label{p1}\ee

%In some sense, this is the reflection that the background
%evolution of collapse dominated by matter is instable.

\textbf{Acknowledgments} We thank R. Brandenberger, Y.F. Cai, Y.
Liu for valuable comments and discussions. This work is supported
in part by NSFC under Grant No:10775180, in part by the Scientific
Research Fund of GUCAS(NO.055101BM03), in part by CAS under Grant
No: KJCX3-SYW-N2.


\begin{thebibliography}{99}

\bibitem{Wands99} D. Wands, Phys. Rev. \textbf{D60}, 023507
(1999).

\bibitem{FB} R. Brandenberger, F. Finelli, JHEP \textbf{0111} (2001) 056;
F. Finelli, R. Brandenberger, Phys. Rev. \textbf{D65}, 103522
(2002).

\bibitem{PP} P. Peter and N. Pinto-Neto, arXiv:0809.2022.

\bibitem{Wands09} D. Wands, arXiv:0809.4556.

\bibitem{Cai0810}  Y.F.Cai, T. Qiu, R. Brandenberger, X.M. Zhang,
arXiv:0810.4677.

\bibitem{S} A.A. Starobinsky, JETP Lett. \textbf{30}, 682 (1979).

\bibitem{Tolman} R.C. Tolman, Relativity, Thermodynamics and Cosmology, (Oxford U. Press,
Clarendon Press, 1934).

\bibitem{STS} P.J. Steinhardt and N. Turok, Science \textbf{296},
(2002) 1436; Phys. Rev. \textbf{D65} 126003 (2002).

\bibitem{LS} J.L. Lehners, P.J. Steinhardt, arXiv:0812.3388.

\bibitem{Xiong} H.H. Xiong, Y.F. Cai, T. Qiu, Y.S. Piao, X.M.
Zhang, Phys. Lett. \textbf{B666}, 212 (2008).

\bibitem{Piao04} Y.S. Piao, Phys. Rev. \textbf{D70}, 101302 (2004);
Y.S. Piao and Y.Z. Zhang, Nucl. Phys. \textbf{B725}, 265 (2005).


\bibitem{BD} J. Barrow and M. P. Dabrowski, Mon. Not. R. Astr. Soc. \textbf{275}, 850
(1995).

\bibitem{KSS} N. Kanekar, V. Sahni and Y. Shtanov, Phys. Rev. \textbf{D63}, 083520
(2001).


\bibitem{Lidsey04} J.E. Lidsey, D.J. Mulryne, N.J. Nunes and R. Tavakol,
Phys. Rev. \textbf{D70}, 063521 (2004).

\bibitem{BF} L. Baum and P.H. Frampton, Phys. Rev. Lett. \textbf{98},
071301 (2007); P.H. Frampton, Mod. Phys. Lett. \textbf{A22}, 2587
(2007).

\bibitem{CB} T. Clifton and J.D. Barrow, Phys. Rev. \textbf{D75}, 043515 (2007).

\bibitem{Xin} X. Zhang, arXiv:0708.1408.

\bibitem{Biswas} T. Biswas, arXiv:0801.1315;  T. Biswas, S.
Alexander, arXiv:0812.3182.

\bibitem{NB} M. Novello, S.E.P. Bergliaffa, Phys. Rept. \textbf{463}, 127 (2008).

\bibitem{Erick} J.K. Erickson, S. Gratton, P.J. Steinhardt, N.
Turok, Phys. Rev. \textbf{D75}, 123507 (2007).

\bibitem{DH} D.H. Lyth, Phys. Lett. \textbf{B524}, 1 (2002);
Phys. Lett. B 526, 173 (2002).


\bibitem{DV} R. Durrer and F. Vernizzi, Phys. Rev. \textbf{D66}, 083503
(2002).

\bibitem{TBF} S. Tsujikawa, R. Brandenberger, F. Finelli,
Phys. Rev. \textbf{D66}, 083513 (2002).

\bibitem{GKS} S. Gratton, J. Khoury, P.J. Steinhardt, N. Turok, Phys. Rev.
\textbf{D69}, 103505 (2004).

\bibitem{Piao0403} Y.S. Piao, Y.Z. Zhang, Phys. Rev. \textbf{D70}
043516 (2004).

\bibitem{BV} V. Bozza, G. Veneziano, JCAP 009, \textbf{005} (2005); V.
Bozza, JCAP 0602, \textbf{009} (2006);

\bibitem{ABB} S. Alexander, T. Biswas and R.H. Brandenberger,
arXiv:0707.4679.

\bibitem{Cai11} Y.F. Cai, T. Qiu, R. Brandenberger, Y.S. Piao and
X.M. Zhang, JCAP 0803, \textbf{013} (2008); Y. F. Cai and X.
Zhang, arXiv:0808.2551.

\bibitem{AW} L.E. Allen, D. Wands, Phys. Rev. \textbf{D70}, 063515
(2004).

\bibitem{HV} J.C. Hwang, E.T. Vishniac, Astrophys. J. \textbf{382}, 363
(1991).

\bibitem{DM} N. Deruelle, V.F. Mukhanov, Phys. Rev. \textbf{D52},
5549 (1995).

\bibitem{83} A. Vilenkin, Phys. Rev. \textbf{D27}, 2848 (1983).

\bibitem{86} A. Linde, Phys. Lett. \textbf{B175}, 395 (1986).

\bibitem{Brandenberger93} R. Brandenberger, V. Mukhanov, T.
Prokopec, Phys. Rev. Lett. \textbf{69} 3606 (1992).

\bibitem{Brandenberger931} R. Brandenberger, T.Prokopec, V. Mukhanov,
Phys. Rev. \textbf{D48} 2443 (1993).

\bibitem{PFZ} Y.S. Piao, B. Feng, X.M. Zhang, Phys. Rev. \textbf{D69},
103520 (2004).

\bibitem{PTZ} Y.S. Piao, S. Tsujikawa, X.M. Zhang, Class. Quant. Grav. \textbf{21}, 4455 (2004).





\end{thebibliography}
\end{document}